\documentclass[12pt]{book}

\usepackage[dvips]{graphicx,color}
\usepackage{makeidx,tsukuba}

\makeauthorindex
\makeindex

\begin{document}

\BookTitle{\itshape The 28th International Cosmic Ray Conference}
\CopyRight{\copyright 2003 by Universal Academy Press, Inc.}
\pagenumbering{arabic}

\chapter{
Analog read-out of the RPCs in the ARGO-YBJ experiment}

\author{
%
%
M. Iacovacci,$^1$ S. Catalanotti,$^1$ P. Creti,$^2$ G. Liguori,$^3$ 
L. Saggese$^1$ for the ARGO-YBJ Collaboration [3]\\
{\it (1) Dip. di Fisica Universit\'a di Napoli and INFN sez. di Napoli, Napoli, 
Italy\\
(2) Dip. di Fisica Universit\'a di Lecce and INFN sez. di Lecce, Lecce, Italy \\
(3) INFN sez. Pavia, Pavia, Italy}\\
}

\section*{Abstract}
The ARGO-YBJ experiment, currently under construction at the Yangbaijing Laboratory (4300 m a.s.l.), consists of a single layer of $\sim 2000$ Resistive Plate Chambers (RPCs) for a total instrumented area of $\sim$ 6700 m$^2$. The digital read-out, performed by means of pick-up electrodes 6.7 $\times$ 62 cm$^2$ ('strips'), allows one to measure the particle number of small size showers. To extend the size range up to the knee region it is necessary to implement the charge read-out of the detector chambers. In order to achieve this goal each RPC has been instrumented with two large size pads of dimensions 140$\times$ 125 cm$^2$. In this paper the performance of the prototype circuit devoted to the charge read-out is reported.

\section{Introduction}

The ARGO-YBJ detector is located in Tibet (P.R. China) at the Yangbajing High Altitude Cosmic Ray Laboratory ($30^{\circ}$.11 N, 90$^{\circ}$.53 E, 4300 m a.s.l., 606 g/cm$^2$).
It consists of a central carpet of a single layer of RPCs operated in streamer mode, 74 $\times$ 78 m$^2$ size, with an active area of $\sim$ 92$\%$, surrounded by a guard ring [3].
\begin{figure}[t]
\vfill   
\vspace{-0.2pc}
\hspace{-0.5pc}
\begin{minipage}[t]{0.50\linewidth}
   \begin{center}
    \includegraphics[height=17pc]{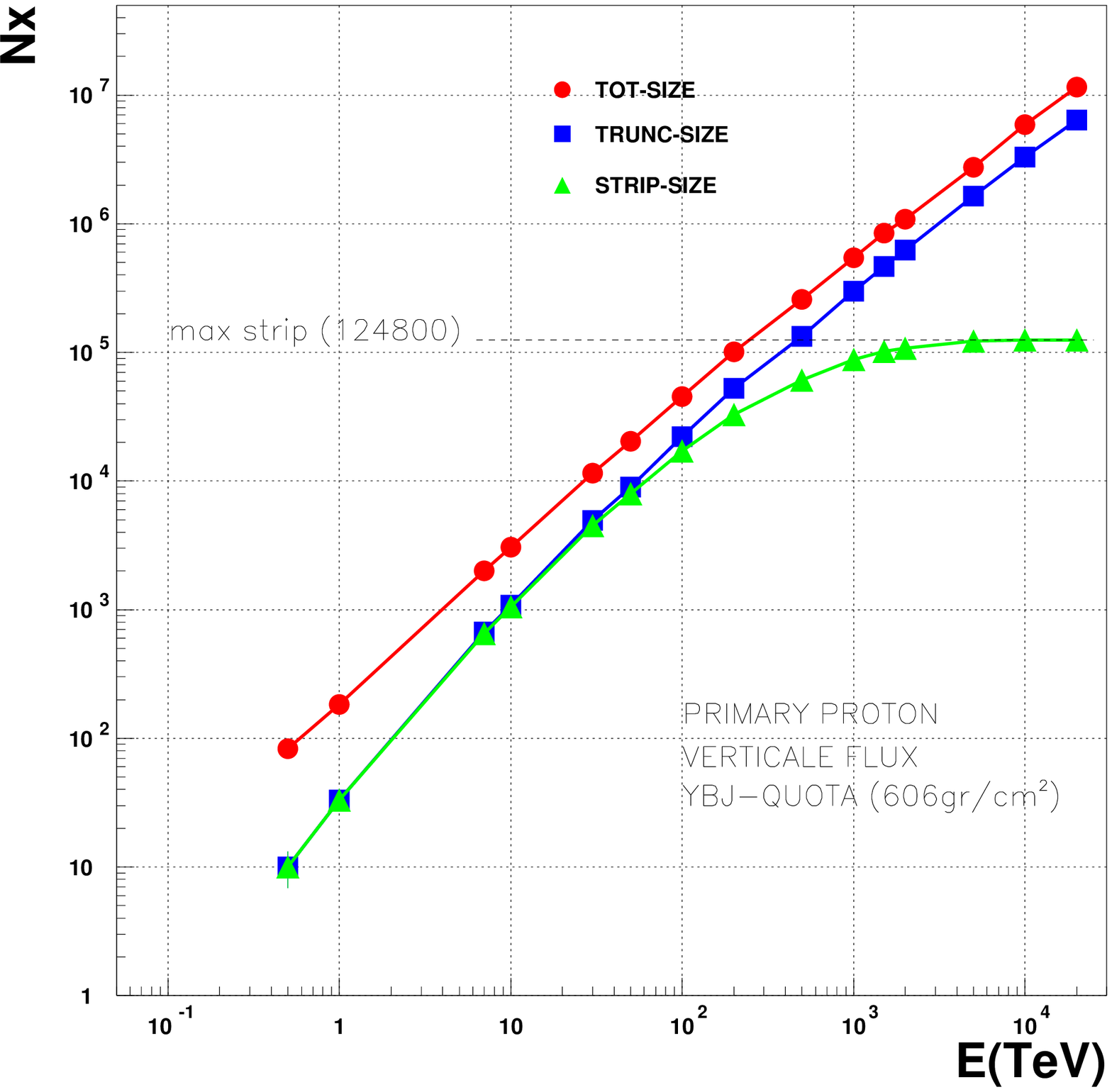}
  \end{center}
  \vspace{-0.4pc}
   \caption{Average strip size compared to the total size and truncated size .}
\end{minipage}\hfill
\hspace{-1.5cm}
\begin{minipage}[t]{.45\linewidth}
  \begin{center}
    \includegraphics[height=22pc]{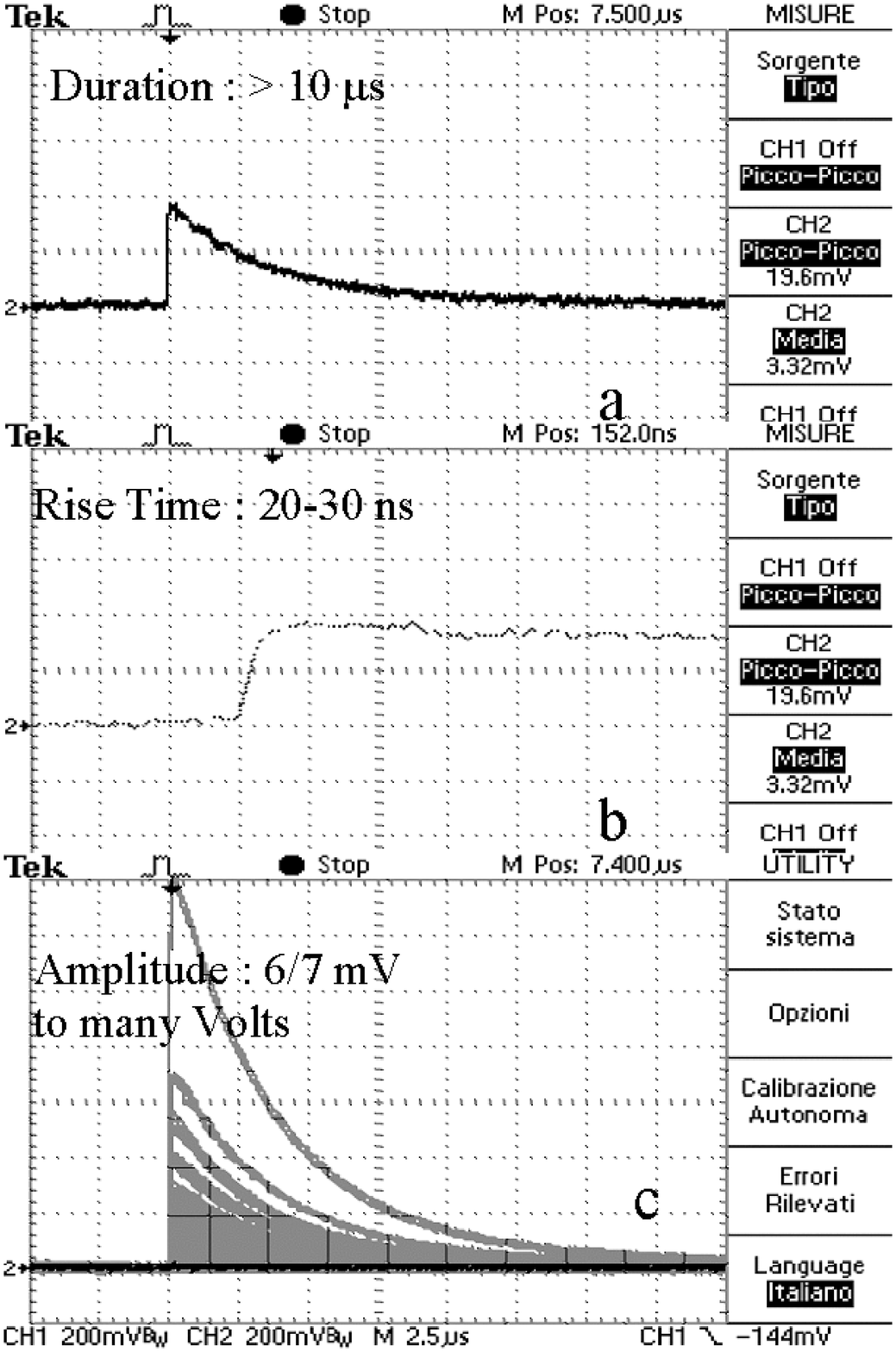}
  \end{center}
  \vspace{-0.5pc}
    \caption{Typical m.i.p. signal from the Big Pad ($50 ~ \Omega$ load, see text) .}
\end{minipage}\hfill
\end{figure}
The signals from each RPC are picked-up with $80$ read-out strips with an average density of $\sim 22~strips/m^2$. A simulation has been carried out by means of the CORSIKA/QGSjet code [1] in order to study the dependence on the energy of the number of fired strips for quasi-vertical ($<15^o$) showers with core in a fiducial area ($A_f$) of $\sim 260~m^2$ at the center of the carpet [2]. The average strip size ($N_s$) is compared in Fig.\,1 to the total size and to the size sampled by the central carpet (truncated size) for proton induced showers. Fig.\,1 shows clearly that the digital response of the detector cannot be used to study the primary spectrum at energies above a few hundred TeV. In order to extend the dynamic range a charge read-out has been implemented by instrumenting each RPC with two large size pads  $140 \times  125~ cm^2$ each (Big Pad). Here we report on the performance of a prototype circuit devoted to the read-out of the Big Pad 
signals.

\section{The Charge Meter Circuit }

The Big Pad is read-out via a unity gain Peak-and-Hold (PH) circuit, specifically designed to fulfill the signal requirements. A typical m.i.p. signal in the Big Pad, with the detector operated at $9.5~ kV$, at sea level, with a gas mixture made of $15\%$ Ar, $10\%$ Isobutane and $75\%$ R143a, is shown in Fig.\,2.  There the same signal is reported in a) with a horizontal scale of $2.5~ \mu s$ and in b) with a horizontal scale of $100~ ns$. In both cases the vertical scale is $10 ~mV$. In Fig.\,2c the vertical scale is $200 ~mV$ (horizontal scale $2.5~ \mu s$) with the oscilloscope set at infinite persistence and triggered for $20'$ by single m.i.p. crossing the chamber.  

\begin{figure}[t]
\hspace{1.pc}
\vfill \begin{minipage}[t]{.47\linewidth}
  \begin{center}
    \includegraphics[height=15pc]{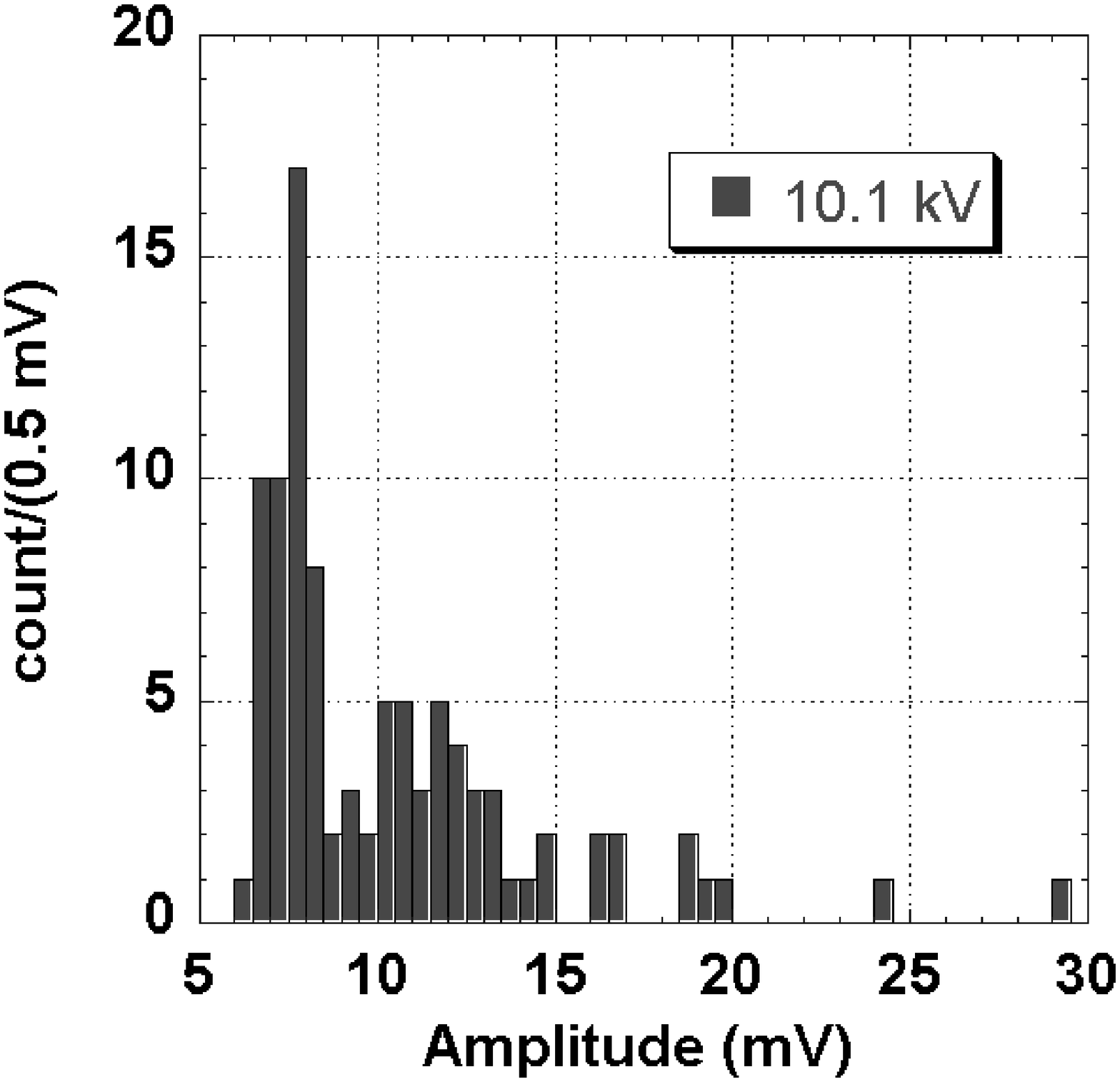}
  \end{center}
  \vspace{-0.3pc}
   \caption{Amplitude distribution of the Big Pad signal at $10.1~kV$ .}
\end{minipage}\hfill
\hspace{-2.5cm}
\begin{minipage}[t]{.5\linewidth}
  \begin{center}
    \includegraphics[height=16pc]{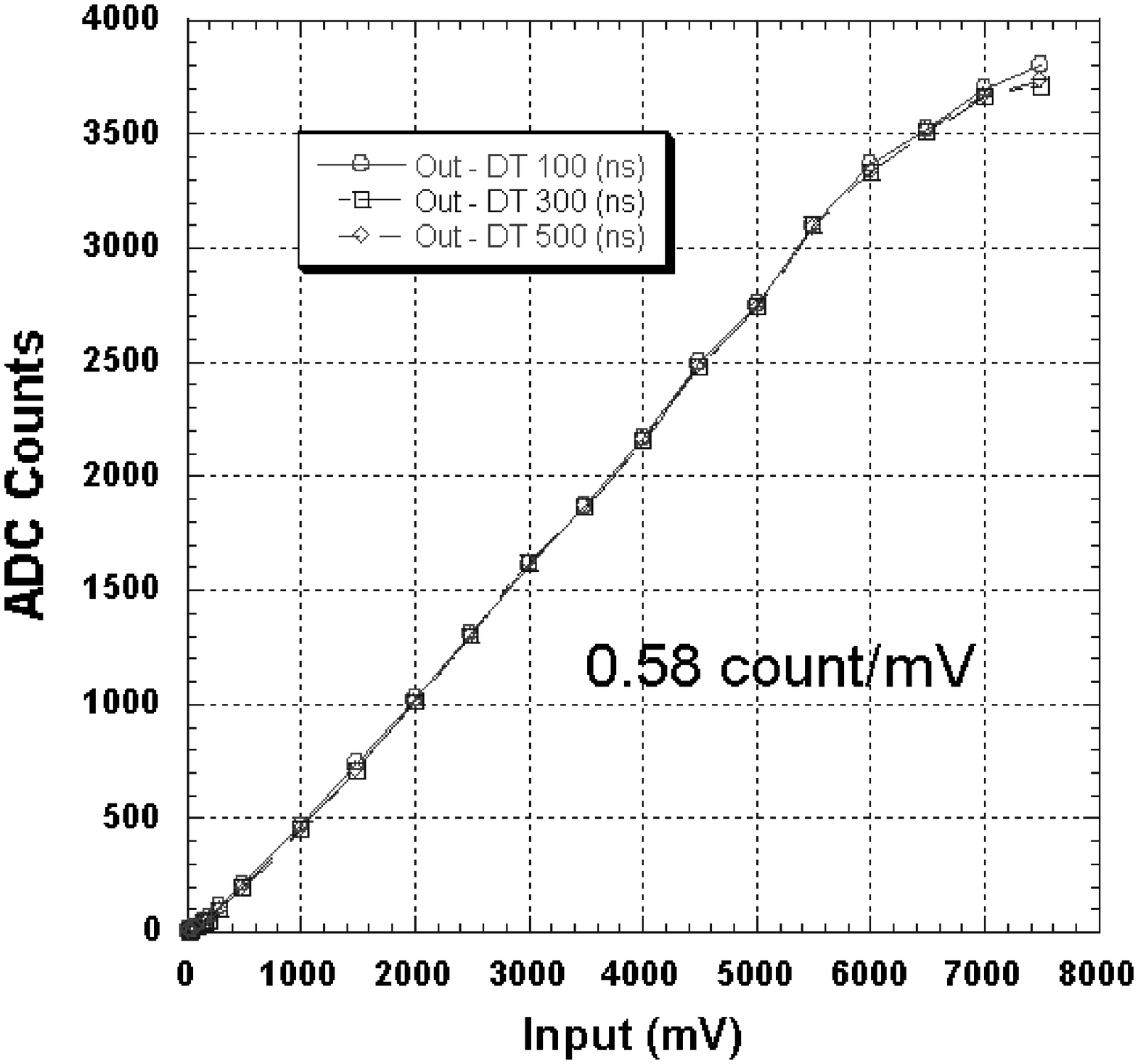}
  \end{center}
  \vspace{-0.5pc}
    \caption{Calibration curve of the charge meter circuit for three different trigger delay time .}
\end{minipage}\hfill
\end{figure}

A single particle pulse has a rise time in the range of a few tens of ns and a discharge time constant of many microseconds; the amplitude of the Big Pad signal, on a $50~ \Omega$ load, can be as high as many Volts as shown in Fig.\,2c. The PH circuit is able to track the Big Pad signals starting from $30~ mV$. When this threshold is exceeded, the PH generates a timing gate of $1.5~\mu$s and in this window it keeps the maximum value of the input signal. The presence of a trigger during this hold-and-update phase validates the event and then PH output level is digitized by a 12-bit ADC with a conversion time of 1.2 $\mu$s. If the gate time elapses and no trigger arrives, the PH restarts the operations and clears its output. 

The Big Pad pulse height allows us to reconstruct the number of incident particles. In Fig.\,3 is reported the amplitude distribution of the m.i.p. signal provided by the Big Pad on $50~ \Omega$, seen at the oscilloscope: there is a first peak at $\sim 6~mV$ with a H.W.H.M. of $\sim 1~mV$, that corresponds to one streamer, and a second peak at about $12~mV$ which is wider and corresponds to two streamers. 

The output discharge rate of the PH introduces an error in the digitized data that depends upon the jitter in the trigger arrival time. In the ARGO-YBJ experiment, a delay in the trigger generation up to $400~ns$ is expected for events at the maximum slant. The PH discharge rate is $\sim 6~ mV/\mu s$, it is almost independent on the signal amplitude, and corresponds in the worst case to an error of one in the particle counting. 

Fig.\,3 shows the calibration curve of the circuit for three different values of the trigger delay time, namely $100$, $300$ and $500~ ns$. The input signal has been provided by a digital pulse generator able to generate signals according to the shape of the signal produced by a m.i.p. on the Big Pad. The figure shows a good linearity from $100~ mV$ up to about $6~ V$. This corresponds to $600 \div 700~particle/m^2$ assuming a mean value of $\sim 6~ mV/m.i.p.$. Above $6~V$ the calibration curve starts to deviate from the linear behaviour, or the circuit response starts to saturate. 

The ADC reading the PH circuit is hosted on a separated board, and both cards are inserted in a VME crate. We verified that the VME power supply does not introduce appreciable noise to the PH circuit neither to the ADC.

\section{Conclusions}

We have realized a test circuit which allows to measure particle density up to $600 \div 700 /m^2$ on our RPCs. The circuit has been tested with signals of many tens of Volts without any breaking. Its response comes out to be linear and independent  on the trigger delay time. Also from the noise point of view it does not present problems. The next step of the analog readout project envisages :
\begin{enumerate}
\item [1.] the extension of the circuit dynamic range up to $10^3 ~ particle/m^2$ at least;

\item [2.]  intercalibration of digital versus analog read-out at low particle density, i.e. $\leq 10~ particle/m^2$;

\item [3.]  realization of a data acquisition system for the analog read-out and its integration with the present, operating digital read-out system at Yangbaijing Laboratory;

\item [4.]  installation of the analog read-out system on a fraction of the ARGO-YBJ apparatus, namely 2 Clusters ($\sim 100 ~m^2$);

\item [5.]  data taking of large size events to study the performance of the system.
\end{enumerate}

\vspace{3.pc}
\section{References}

\vspace{\baselineskip}

\re
1.\ Heck D.\ et al.\ 1998, Report {\bf FZKA 6019} Forschungszentrum Karls\-ruhe.
\re
2.\ Saggese L.\ et al.\ 2003, in this proceedings.
\re
3.\ Surdo A.\ et al.\ 2003, in this proceedings.

\endofpaper
\end{document}